# Negative Refraction with High Transmission in Graphene-hBN Hyper Crystal

Ayed Al Sayem[1]*, M.R.C.Mahdy[2]*, Ifat Jahangir[3], Md. Saifur Rahman[1]

[1]Dept. of EEE, Bangladesh University of Engineering and Technology, Dhaka, Bangladesh

[2] Department of Electrical and Computer Engineering, National University of Singapore, 4 Engineering Drive 3, Singapore

[3]Department of Electrical Engineering, University of South Carolina, Columbia, SC 29208, USA

*Corresponding Authors' Emails: A0107276@nus.edu.sg

And ayedalsayem143@gmail.com

Abstract: **In this article, we have theoretically investigated the performance of graphene-hBN hyper crystals to demonstrate all angle negative refraction. hBN, the latest natural hyperbolic material, can be a very strong contender to form a hyper crystal with graphene due to its excellence as a graphene-compatible substrate. Although bare hBN can exhibit negative refraction, the transmission is generally low due to its high reflectivity. On the other hand, due to graphene's two dimensional nature and metallic characteristics in the frequency range where hBN behaves as a type-I hyperbolic metamaterial, we have found that graphene-hBN hyper-crystals exhibit all angle negative refraction with superior transmission. This has been possible because of the strong suppression of reflection from the hyper-crystal without any adverse effect on the negative refraction properties.** This finding can prove very useful in applications such as super-lensing, routing and imaging within a particular frequency range. We have also presented an effective medium description of the hyper crystal in the low *k* limit and validated the proposed theory using general 4×4 transfer matrix method and also full wave simulation.**

# I. Introduction

Negative refraction [1, 2], hyperbolic metamaterials (HMMs) [3-8] and hyper-crystals [9, 10] have recently developed a keen interest in the field of photonics and nano-photonics. Hyperbolic metamaterials are relatively old compared to the new concept of hyper-crystals [9, 10], which combines properties of hyperbolic metamaterials and photonic crystals and can find very intriguing photonics applications. Hyper-crystals can be made from periodic arrangements of metal and HMM, dielectric and HMM or two different HMMs [9]. In HMM, (the core of hyper-crystal) the components of the permittivity tensor have opposite signs in two orthogonal directions and so unbounded high k bulk propagating waves are supported in HMMs [3-8]. Negative refraction is a very common phenomenon in HMMs and has been demonstrated in visible [11], mid infrared [12] and THz ranges [13] for TM polarized electromagnetic waves. Negative refraction has also been demonstrated in uniaxial magnetic hyperbolic ferro materials [14] for TE polarized electromagnetic waves. Not only that, polar dielectrics [15], as well as graphene based hyperbolic metamaterials [13], have also been investigated for negative refraction. HMMs are generally made using alternate metal dielectric multi-layers [3, 7, 8], metallic nano-rods in a dielectric host [16-18] or they can be hyperbolic naturally such as hBN [19-27], which is the latest addition to the hyperbolic metamaterial family [19-27]. Since a hyper-crystal contains multiple layers of HMMs, natural materials that exhibit HMM behaviors by themselves are generally the preferred choice [9, 10] for constructing such hyper crystals. Graphene, on the other hand, has found enormous applications in the recent years in both photonics and plasmonics in a broad frequency ([27, 28], mid infrared [29, 30] to visible) range because of its unique and tunable optical properties. Recently, graphene has also been used in constructing HMMs [31-33]. Because of having similar (hexagonal) crystal structure, hBN is widely used a substrate material for graphene as hBN provides an amazing clean environment. And because of this, hBN insulator is promised to be most suitable substrate for van der Waals hetero-structures [34, 35]. These facts invoke a natural curiosity about the properties and possible applications of graphene-hBN hyper crystals.

In this article, we theoretically investigate the properties of graphene-hBN hyper crystals (cf. Fig. 1), especially the one associated with negative refraction. We have also presented an effective medium theory based description of the hyper crystal in the low k limit. It has been found that graphene-hBN hyper crystals can exhibit all angle negative refraction with very high transmission.While bare hBN can also demonstrate negative refraction; transmission turns out to be low due to high reflectivity. But using graphene-hBN hyper-crystals, capitalizing on graphene's true two dimensional nature and metallic characteristics in the particular frequency range where hBN shows type-I HMM behavior; it is possible to strongly suppress reflection from the hyper-crystal without any adverse effect on the negative refraction properties. We believe this work will be very useful for future investigations on intriguing properties and applications of graphene-hBN based hyper crystals; especially negative refraction with high transmission, which may find practical applications such as super-lensing, routing and imaging in that particular frequency range of interest.

## II.   Theory

We have organized the theory in three sub-sections. Sub-section A describes the optical properties of graphene and hBN in the frequency range of interest. Sub-section B describes the basic theory of hyper crystal and effective medium approximation and its applicability and validity in the particular wave vector region. This effective medium description is very helpful to easily understand the critical role played by graphene in the hyper-crystal and its high transmission. Sub-section C describes the all-angle negative refraction properties of graphene-hBN hyper crystal and its transmission, reflection and absorption by numerical calculations along with comparisons made using commercial simulation packages.

## A. Graphene, hBN Optical Properties

In this sub-section, we briefly present the optical properties of graphene, which will be used throughout the article.

Dependence of optical conductivity of graphene on chemical potential, temperature, frequency and relaxation time can be determined using the Kubo formalisms [36, 37] including both intra-band and inter-band contributions,

$$\sigma(\omega,\mu_c,\tau,T) = \frac{e^2}{\pi\hbar^2}\frac{i}{\omega-i\tau^{-1}}\left[\frac{1}{(\omega-i\tau^{-1})^2}\int_0^\infty \xi\left(\frac{\delta f_d(\xi)}{\delta\xi}-\frac{\delta f_d(-\xi)}{\delta\xi}\right)\delta\xi - \int_0^\infty \frac{f_d(-\xi)-f_d(\xi)}{(\omega-i\tau^{-1})^2-4(\xi\hbar^{-1})^2}\delta\xi\right] - \quad (1)$$

The first and second part of equation (1) corresponds to the intra-band and inter-band contribution respectively. The intra-band contribution can be simplified as,

$$\sigma_{intra}(\omega,\mu_c,\tau,T) = \frac{-ie^2 K_B T}{\pi\hbar^2(\omega-i\tau^{-1})}\left[\frac{\mu_c}{k_B T}+2\ln\left(e^{-\frac{\mu_c}{k_B T}}+1\right)\right] \quad (2)$$

And the inter-band contribution can be approximated as ($\mu_c \gg k_B T$),

$$\sigma_{inter}(\omega,\mu_c,\tau,T) = \frac{-ie^2}{4\pi\hbar}\ln\left(\frac{2|\mu_c|-(\omega-i\tau^{-1})\hbar}{2|\mu_c|+(\omega-i\tau^{-1})\hbar}\right) \quad (3)$$

So the total conductivity Graphene can be given by,

$$\sigma(\omega,\mu_c,\tau,T) = \sigma_{intra}(\omega,\mu_c,\tau,T) + \sigma_{inter}(\omega,\mu_c,\tau,T) \quad (4)$$

Where, $\omega$ is the angular frequency, $\mu_c$ is the chemical potential, $\tau$ is the relaxation time, e is the electron charge, $\hbar$ is the reduced plank's constant, $k_B$ is the Boltzmann's constant, T is the temperature, $\xi$ is energy and $f_d(\xi)$ is the Fermi Dirac distribution given by, $f_d(\xi) = $

$\left(1 + e^{\frac{\xi - \mu_c}{K_B T}}\right)^{-1}$ Because of its 2D nature, graphene is basically an optically uni-axial anisotropic material, whose permittivity tensor can be given by,

$$\varepsilon_{graphene} = \begin{bmatrix} \varepsilon_{g,t} & 0 & 0 \\ 0 & \varepsilon_{g,t} & 0 \\ 0 & 0 & \varepsilon_{g,\perp} \end{bmatrix} \quad (5)$$

Graphene's tangential permittivity can be given by,

$$\varepsilon_{g,t} = 1 + j \frac{\sigma(\omega, \mu_c, \tau, T)}{\omega \varepsilon_o t_g} \quad (6)$$

Where, $\omega$ is the angular frequency, $\varepsilon_o$ is the free space permittivity and $t_g$ is the thickness of monolayer graphene. As graphene is a two dimensional material, the normal electric field cannot excite any current in the graphene sheet, so the normal component of the permittivity is given by $\varepsilon_{g,n} = 1$ [38]. For few layer graphene, the tangential or in plane permittivity (equation (7)) is the same as mono layer graphene [39, 40]. However, for few layer graphene, normal component of permittivity $\varepsilon_{g,\perp}$ can be approximated as graphite's permittivity (~2).

Considering only single-Lorentzian form, the principal dielectric tensor components of hBN can be expressed as [24, 41, 42],

$$\varepsilon_{uu} = \varepsilon_\infty \left(1 + \frac{(\omega_{LO,u})^2 - (\omega_{TO,u})^2}{(\omega_{TO,u})^2 - \omega^2 - i\omega \gamma_u}\right) \quad (7)$$

Where $u = x, y$ represents the transverse (a, b crystal plane) and $u = z$ represents the z (c crystal axis) axes. LO, TO, $\varepsilon_\infty$ and $\gamma$ represents the LO and TO phonon frequencies, the high-frequency dielectric permittivity and the damping constant respectively. These data has been taken from [24]. In the frequency range where $\omega$ lies between LO and TO frequency, $\varepsilon_{uu}$ becomes negative.

Fig. 2a shows the real and imaginary part of the parallel /tangential permittivity of graphene as a function of wave-number for various chemical potential values. For higher chemical potential values tangential permittivity becomes more and more negative. It should be noted that in this spectral region intra-band transition in graphene dominates and loss is very low due to Pauli blocking of inter-band transition [43]. In this spectral region, the loss arises due to imperfection or defects in graphene and so for a finite scattering time.

Fig. 1 shows the schematic of the proposed graphene-hBN hyper crystal. Fig. 2b and 2c shows the real and imaginary part of the calculated tensor components respectively in the frequency range from 700 to 1800 cm$^{-1}$. hBN is a layered van der Waals crystal and the ab- and c axis-polarized phonons are separated in frequency by a great margin. As a result, two Reststrahlen bands are non-overlapping ([19], [24-26]). Due to this non-overlapping Reststrahlen bands, both Type I and Type II hyperbolic responses are available in hBN. In the lower Reststrahlen band, $(\omega_{LO} = 825 \text{cm}^{-1}, \omega_{TO} = 760 cm^{-1})$ $\varepsilon_\parallel > 0$ and $\varepsilon_\perp < 0$, and thus leading to the Type I hyperbolic response with bandwidth of 1.04 μm. In the upper Reststrahlen band, $(\omega_{LO} = 1614 \text{cm}^{-1}, \omega_{TO} = 1360 \text{cm}^{-1})$ $\varepsilon_\parallel < 0$ and $\varepsilon_\perp > 0$ and thus leading to the Type II hyperbolic response with bandwidth of 1.16 μm

## B. Hyper Crystal theory and Effective medium description

For a general one dimensional anisotropic crystal or hyper-crystal where $d_1$ and $d_2$ are the thickness of the layers and each having the permittivity tensor of the form,

$$\varepsilon_i = \begin{bmatrix} \varepsilon_{xi} & 0 & 0 \\ 0 & \varepsilon_{yi} & 0 \\ 0 & 0 & \varepsilon_{zi} \end{bmatrix} ; i = 1,2 \quad (8)$$

The basic Bloch equation for p polarized wave for the anisotropic crystal have been derived as (cf. Supplementary for mathematical derivation),

$$\cos(k_y d) = \cos(k_{y1} d_1) \cos(k_{y2} d_2) - \frac{1}{2}\left(\frac{\varepsilon_{x2} k_{y1}}{\varepsilon_{x1} k_{y2}} + \frac{\varepsilon_{x1} k_{y2}}{\varepsilon_{x2} k_{y1}}\right) \sin(k_{y1} d_1) \sin(k_{y2} d_2) \quad (9)$$

Where, $d = d_1 + d_2$ and $k_{y1}$ and $k_{y2}$ are the propagating wave vectors in layer 1 and 2 respectively and can be given by,

$$k_{yi} = \sqrt{\varepsilon_{xi} k_o^2 - \frac{\varepsilon_{xi}}{\varepsilon_{yi}} k_x^2} \ ; \ i = 1,2 \quad (10)$$

Where, $k_o$ is the wave vector in free-space. Now, for isotropic case where both layers have isotropic permittivity (for example, metal-dielectric multilayers) $k_o d \ll 1$ is the sufficient condition for effective medium theory. But for hyper-crystal, the situation is a little different [9] because either or both layers in the unit cell now can support high $k > k_o$ waves. So, effective medium theory cannot properly describe the physical properties in hyper crystal, such as the band gap originating from high k waves in either or both layers. But for low k waves ($k < k_o$) and for moderate high k waves, effective medium theory should still work, where layer thickness $d_1$, $d_2$ and wavelength $\lambda$ plays a critical role. Effective permittivity tensor for hyper-crystal can thus be derived from equation (10) and they are given by (cf. Supplementary for mathematical derivation),

$$\varepsilon_{xeff} = f\varepsilon_{x1} + (1-f)\varepsilon_{x2} \quad (11a)$$

$$\varepsilon_{yeff} = \left(\frac{f}{\varepsilon_{y1}} + \frac{1-f}{\varepsilon_{y2}}\right)^{-1} \quad (11b)$$

And from equation (9) and (11a, 11b) the wave vector in the propagating direction for the hyper-crystal, $k_{y,hc}$ can be given by,

$$k_{y,hc} = \sqrt{\varepsilon_{xeff} k_o^2 - \frac{\varepsilon_{xeff}}{\varepsilon_{yeff}} k_x^2} \qquad (12)$$

In the following discussion, we demonstrate whether effective medium theory works or to what extent if it does. Here we do not put any arbitrary tensor values in equation (8), instead we consider graphene and hBN as layer 1 and layer 2, as they are the primary topic of interest in this article.

Fig. 3 shows the schematic representations of the equi-frequency contour in the wave vector space for the graphene-hBN hyper-crystal (within the loss less limit) at $\omega_c = 800 cm^{-1}$(a, b), at $\omega_c = 1600 cm^{-1}$(c, d). The red symbols represent the actual calculation (equation (9)) and effective medium approximation (equation (12)). At $\omega_c = 800 cm^{-1}$ (in the lower Reststrahlen band) and $\omega_c = 1600 cm^{-1}$(in the upper Reststrahlenband), hBN behaves as a type I and type II hyperbolic metamaterial respectively. It can be clearly visualized that effective medium theory fails to describe the propagation behavior and band gap for the high k waves. Figure 3c and 3d represents the close-up view of the iso-frequency contour presented in 3a and 3b respectively. It can be clearly observed that effective medium theory can adequately represent the propagation behavior at low k and moderate high k waves.

## C. Negative Refraction in Graphene-hBN Hyper-Crystal

From the (hyperbolic) dispersion curves in Fig. 3, it is clearly visible that negative refraction will also be available in hyper-crystals. Here, we consider only the lower Reststrahlen band, where hBN behaves as a Type-I hyperbolic media. Angle of refraction for power flow (group velocity) and wave vector can be calculated as in [10] but as we can use effective medium approximation, we can have a simplified equation as discussed below.

Refraction angles for the wave vector in a uniaxial anisotropic medium can be given by [11],

$$\theta_{r,hc} = \tan^{-1}\left(\frac{k_x}{k_{y,hc}}\right) \quad [13a]$$

And refraction angles for the power flow or group velocity can be given by,

$$\theta_{s,hc} = \tan^{-1}\left(\frac{k_x \varepsilon_{xeff}}{k_{y,hc} \varepsilon_{yeff}}\right) \quad [13b]$$

The refractive group index for the hyper crystal can thus be given by,

$$n_{g,hc} = \frac{n_{air} \sin(\theta_{inc})}{\sin(\theta_s)} \quad [14]$$

Where $\theta_{inc}$ is the incident angle. Fig. 4a shows the refracted angles for the pointing vector (symbols) and wave vector (solid lines) as function of incident angle at wavenumber 810 cm$^{-1}$ for graphene-hBN hyper crystal for different values of chemical potential of graphene. Fig. 4b shows the 2D map of the refracted angles for the pointing vector as a function of wavenumber and incident angle. Fig. 4a and 4b clearly demonstrate that in the lower Reststrahlen band ($760 cm^{-1} \leq \omega \leq 825 cm^{-1}$), graphene-hBN hyper crystal exhibit all angle negative refraction.

Fig. 5a shows (red curve) the calculated transmission, reflection and absorption as a function of frequency for an hBN slab of thickness 1 μm. In the lower Reststrahlen band, in spite of low absorption, transmission is relatively moderate because of the high reflection due to impedance mismatch (high in plane permittivity of hBN). The same figure (symbols for effective medium approximation based calculation and solid lines for general transfer matrix method [44-46]) shows the transmission, reflection and absorption from a hyper-crystal made from graphene and hBN (5 layers of alternating graphene and hBN layers where thickness of each hBN layer is 200nm) with total thickness of ~1um. It can be clearly observed that Graphene hBN hyper crystal shows better transmission with higher chemical potential of graphene (both bare hBN and hyper crystal have same thickness). Besides, both effective medium theory and general transfer matrix calculation show similar results.

The physics behind higher transmission can be easily understood. Reflection of bare hBN arises due to the higher dielectric tangential permittivity. But in graphene-hBN hyper crystal, the tangential permittivity is the effective permittivity of graphene and hBN. And because of graphene's high negative permittivity (cf. Fig. 2a)), effective tangential permittivity decreases from that of the bare hBN (cf. Fig. 2d)) and becomes comparable to the ambient (air) permittivity. As a result momentum mismatch or impedance mismatch reduces and so reflection gets highly suppressed.

It is mentioned earlier that graphene's perpendicular permittivity is small (~2) and also not dispersive. Therefore it has little effect on the perpendicular permittivity of the graphene-hBN hyper crystal, which is required for negative refraction (cf. equation (13b) $\varepsilon_{xeff}$ is positive, so to get negative $\theta_{s,hc}$, one needs $\varepsilon_{yeff} < 0$ ).

Fig 5b shows a 2D map of transmission coefficient as a function of wave-number and incident angle for bare hBN, whereas Fig. 5c and 5d shows the same for graphene-hBN hyper crystal calculated by general 4×4 transfer matrix method [44-46] and effective medium theory respectively. We can clearly observe the complete validation of effective medium approximation for the low k waves.

Fig. 6a shows the transverse magnetic field profile ($H_z$) of a p polarized wave incident from air to hBN and graphene hBN hyper crystal (calculated using full wave simulation, Comsol Multiphysics 4.3b). Fig. 6a is for hBN and Fig. 6b, 6c and 6d are for graphene-hBN hyper crystal with different chemical potential of graphene. As described earlier, with higher chemical potential, graphene's tangential permittivity becomes more negative and tangential permittivity of the graphene-hBN hyper crystal gets closer to the ambient medium (air) permittivity. As a result, lower reflection and higher transmission are achieved without hampering negative refraction. This is clearly illustrated from the full wave numerical simulation in Fig. 6a-6d.

### III. Conclusion

As the research in metamaterial field [47-49] has grown to a saturated state, natural materials are being discovered with exciting properties which was thought to be possessed only by artificial materials or metamaterials. With the discovery of hBN as natural hyperbolic metamaterial, it is obvious that along with new intriguing physics and applications, it will be utilized in almost every possible applications of hyperbolic metamaterials [19-26] because of its low damping and unbounded high k waves. One of the best possible uses of hBN might be in constructing hyper crystals which make a bridge between photonic crystals and hyperbolic metamaterials. In this article, we investigated the properties of graphene-hBN based hyper crystal along with another wonder material graphene [50, 51]. We have found that all angle negative refraction will be available in such hyper crystal with very high transmission with respect to bare hBN. Transmission and negative refraction properties can be controlled by tuning the chemical potential of graphene. We have also given an effective medium description of the hyper crystal in low k limit which might be useful in theoretical calculations and it has been validated by multiple numerical models. We believe that this work could pave the way to the discovery of many intriguing physical phenomena of graphene- hBN hyper crystal and enable many interesting applications to emerge within a very short time.

**Supplementary article:** Supplementary online article contains detail derivations and the details of the numerical calculations.

**Figures and Captions**

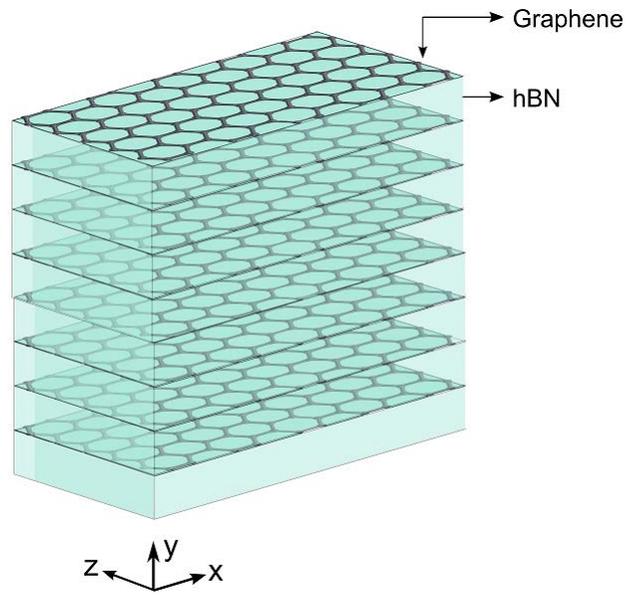

Fig. 1: Schematic of the graphene hBN Hyper Crystal

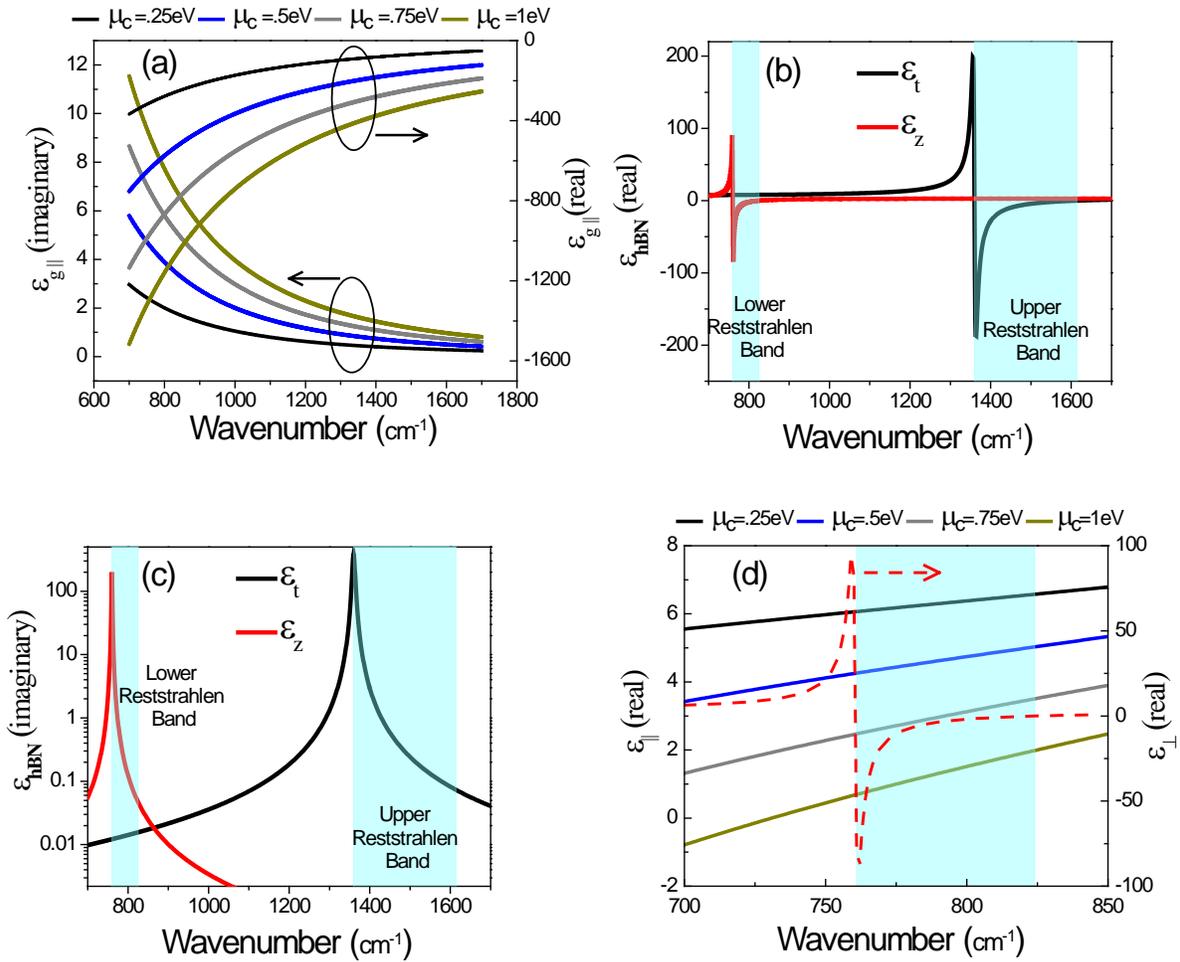

Fig. 2: (a) Real and imaginary part of the tangential $(\varepsilon_{||})$ permittivity of graphene as a function of wavenumber for various chemical potential of graphene. Real (b) and imaginary (c) part of the tangential $(\varepsilon_{||})$ and perpendicular $(\varepsilon_{\perp})$ permittivity of hexagonal boron nitrate. Shaded regions represent the lower and upper Reststrahlen bands. (d) Real and imaginary part of the tangential $(\varepsilon_{||})$ and perpendicular $(\varepsilon_{\perp})$ permittivity of graphene-hBN hyper-crystal. Shaded regions represent the lower Reststrahlen band.

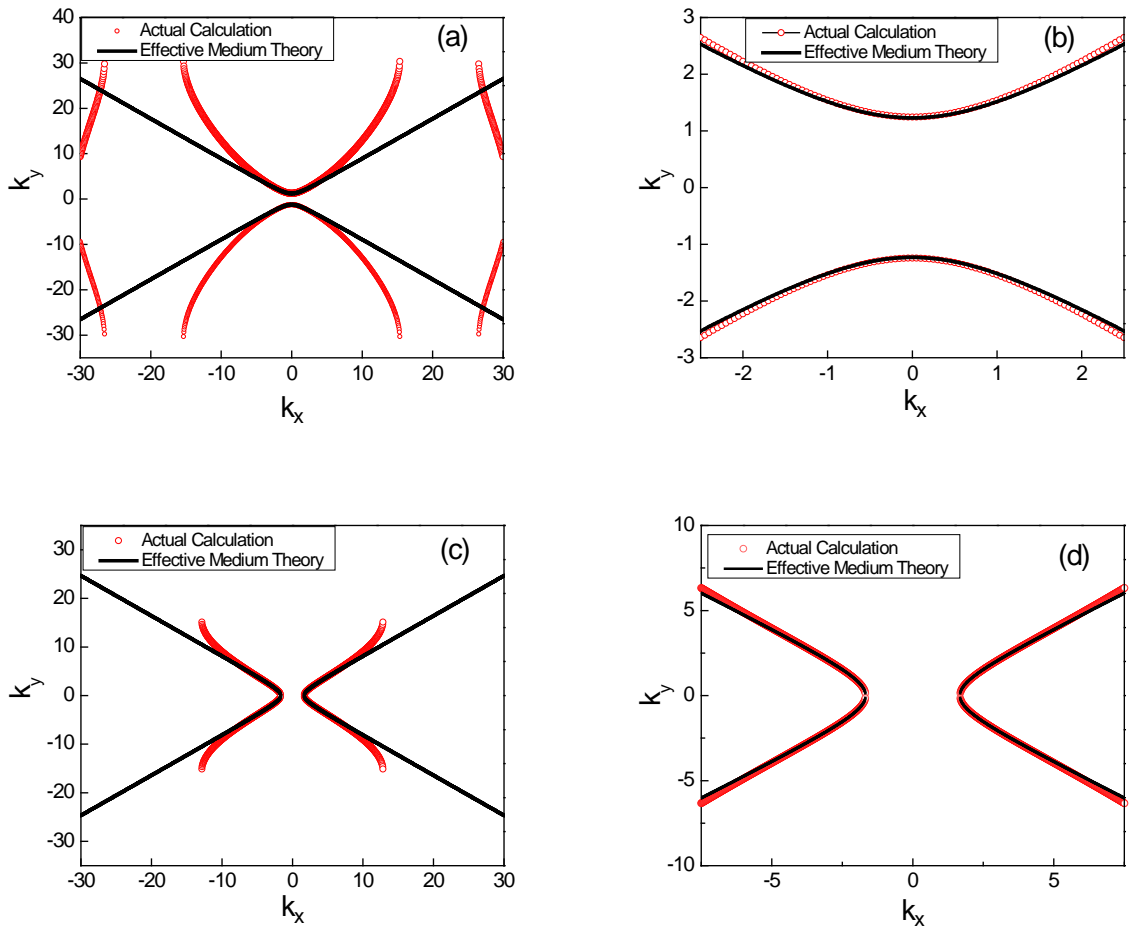

Fig. 3: Iso-frequency Contour of graphene hBN hyper crystal (a) at wavenumber 800cm$^{-1}$ where hBN behaves as Type II hyperbolic metamaterial. (b) at wavenumber 1600cm$^{-1}$ where hBN behaves as Type I hyperbolic metamateria.l (c)-(d) zoom in view of (a) and (b) respectively.

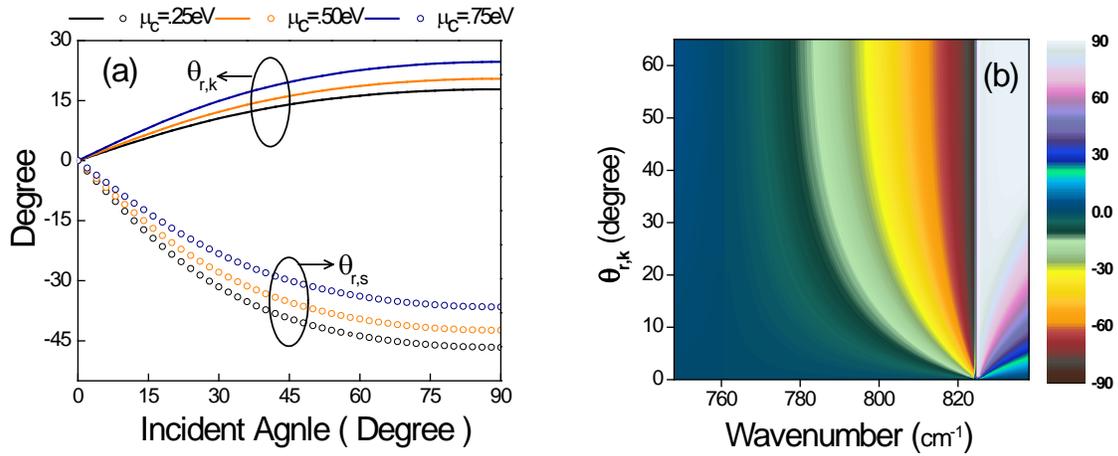

Fig. 4: (a) Refracted angles for the pointing vector (symbols) and wave vector (solid lines) as function of incident angle at wavenumber 810 cm$^{-1}$ for different values of chemical potential of graphene. (b) 2D map of the refracted angles for the pointing vector as a function of wavenumber and incident angle of graphene-hBN hyper-crystal ($\mu_c = .75 eV$).

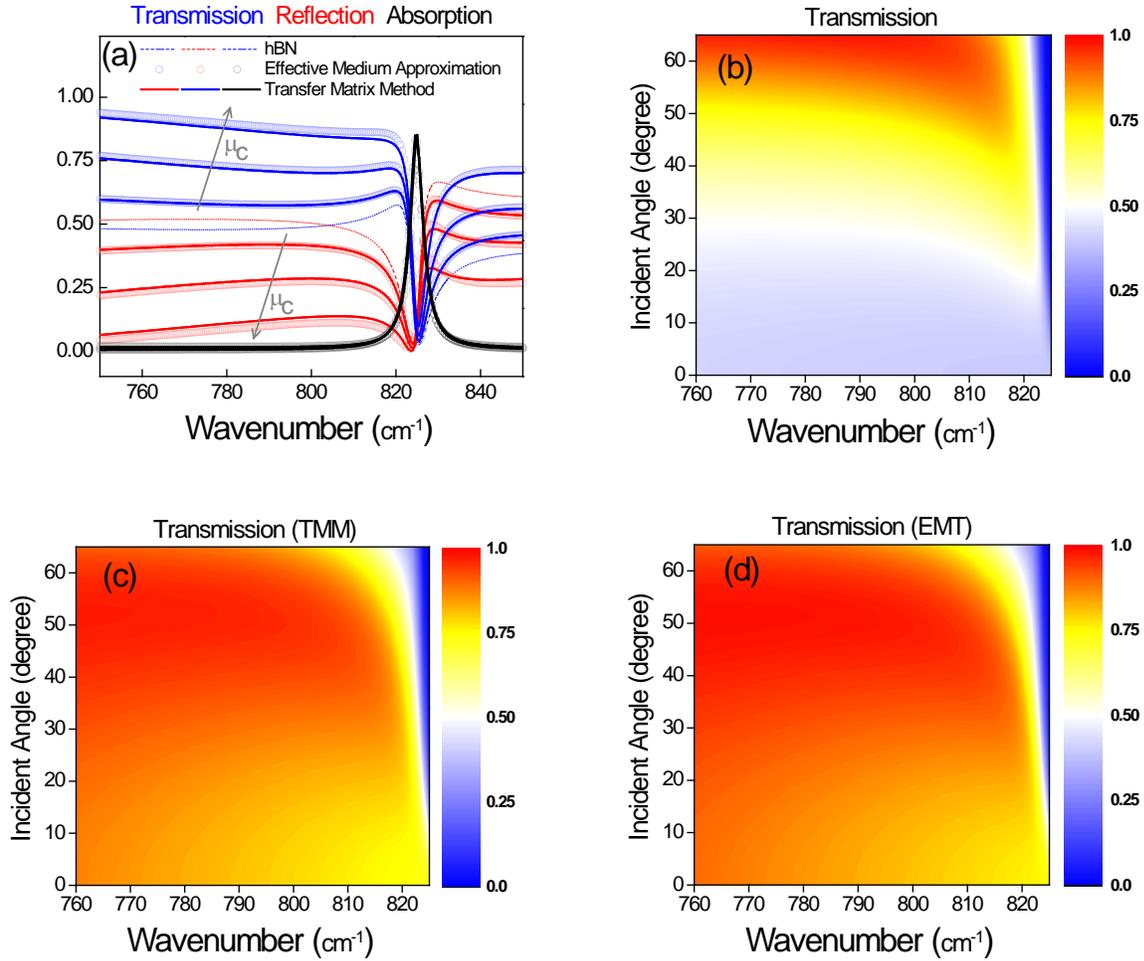

Fig. 5: (a) Transmission (blue), reflection (red) and absorption (black) from hBN and graphene-hBN hyper-crystal as a function of wavenumber. The dotted lines represent calculated values of hBN. The solid lines and symbols represent calculated values of hyper-crystal by general transfer matrix method and effective medium approximation respectively for different chemical potential of graphene (.25eV, .5eV, .75eV). (b), (c) and (d) 2D map of transmission of hBN, graphene-hBN hyper-crystal by general transfer matrix method and effective medium approximation as a function of wavenumber and incident angle.

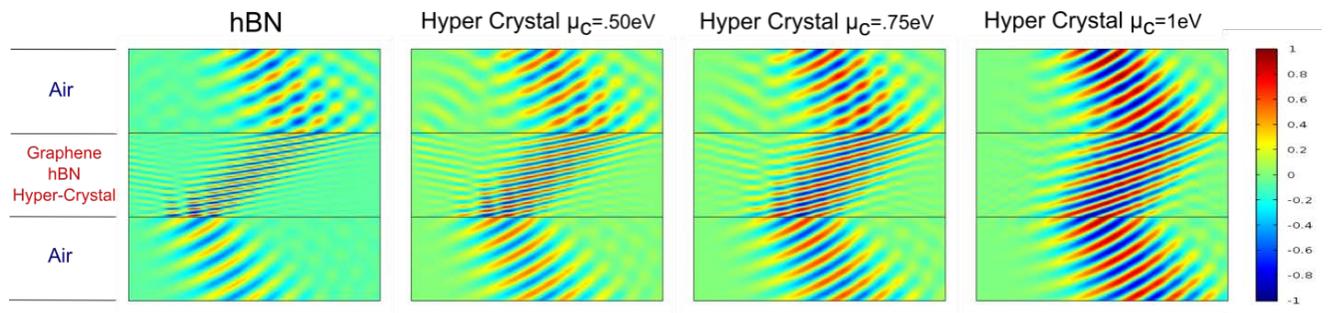

Fig. 6: Full wave simulations (Comsol Multi-physics) (Transverse magnetic field component, Hz) demonstrating negative refraction of a monochromatic TM-polarized Gaussian beam incident from air to (a) hBN and graphene-hBN hyper Crystal (b) $\mu_c = .5eV$ (c) $\mu_c = .75eV$ (d) $\mu_c = 1eV$.